\documentclass{article}
\usepackage{epsfig,amsmath,cite,fullpage,afterpage}

\begin{document}

\begin{center}
\Large Effect of Hydrogen Switchable Mirrors on the Casimir Force
\end{center}

\begin{center}
\large
Davide Iannuzzi, Mariangela Lisanti, and Federico Capasso\footnote{Corresponding author: capasso@deas.harvard.edu}
\end{center}

\begin{center}
\textit{Harvard University, Division of Engineering and Applied Sciences}
\end{center}
\begin{center}
\textit{29 Oxford St., Cambridge, MA - 02138}
\end{center}

\vspace{20pt}

\begin{center}
\textbf{Abstract}
\end{center}
We present systematic measurements of the Casimir force between a gold-coated plate and a sphere coated with a Hydrogen Switchable Mirror (HSM). HSMs are shiny metals that can become transparent by hydrogenation. In spite of such a dramatic change of the optical properties of the sphere, we did not observe any significant decrease of the Casimir force after filling the experimental apparatus with hydrogen. This counterintuitive result can be explained by the Lifshitz theory that describes the Casimir attraction between metallic and dielectric materials.

\vspace{30pt} 
\section{Introduction}

One of the most spectacular consequences of quantum electrodynamics is that it does not leave room for an empty vacuum. Even in the absence of electromagnetic sources, quantum fluctuations of electric and magnetic fields give rise to a {\it zero-point energy} that never vanishes\cite{milonni}.

At first one might think that the zero-point energy is only a constant background to every experimental situation, and, as such, that it has no observable consequences. On the contrary, there are several phenomena in which quantum fluctuations of the electromagnetic field play a very important role, such as the Lamb shift, the anomalous magnetic moment of the electron, spontaneous emission, and the Casimir effect. The latter has received a lot of attention since 1948, when H. B. G. Casimir predicted an attractive force between two perfectly conducting, electrically neutral parallel plates\cite{casimir}. The plates act as a cavity where only the electromagnetic modes that have nodes on both the walls can exist. The zero-point energy when the plates are kept at a distance $d$ is thus smaller than the zero-point energy of free space. Following Casimir's calculation, it is possible to show that, when the plates are brought from infinity to a distance $d$, the energy decreases by $U(d)=E(d)-E(\infty)=-{\pi^2 \hbar c A \over 720 d^3}$, where $A$ is the area of the plates, and $\hbar$ and $c$ are the usual fundamental constants. The Casimir force is thus given by:  

\begin{equation}
F=-{\pi^2 \hbar c A \over 240 d^4} 
\label{casimirclass}
\end{equation} 

\noindent It is worth mentioning that the attraction, though negligible at macroscopic distances, increases rapidly as the distance between the plates decreases. At $d\simeq 100$ nm, the Casimir pressure is as high as $\simeq 10$ ${N\over m^2}$. For this reason, the Casimir effect recently received considerable attention in the context of the development of Micro- and NanoElectroMechanical Systems (MEMS and NEMS) (see for example \cite{serry,hobun1,hobun2}). 

Casimir's theory was generalized to the case of dielectrics by E. M. Lifshitz\cite{lifshitz}, who derived an analytical expression for the attraction between two uncharged parallel plates with arbitrary dielectric functions. The calculated force reduces to the Casimir result (equation \ref{casimirclass}) in the limit of ideal metals (i.e., with infinite plasma frequency). 

The literature on the Casimir force experiments spans almost 50 years, covering the results of a series of measurements of increasing precision \cite{hobun1,hobun2,over,lamo,moh1,moh2,moh3,ederth,bressi,decca1,decca2}. (For a description of the earliest experiments, see for example references \cite{repmoh} and \cite{israe}. See also \cite{iannuzzi,newmost} for a discussion on the precision of some of the most recent measurements). To avoid the problem of keeping two parallel plates at short distances, all the experiments, with the exception of \cite{bressi}, were performed using simpler geometrical configurations, such as a sphere and a plate, or two crossed cylinders. In these cases, if the distance $d$ between the surfaces is much smaller then their radii of curvature $R_{1,2}$ the expected force can be calculated by means of the {\it Derjaguin approximation}\cite{milonni,israe}:

\begin{equation}
F(d)=2\pi\times \bigl( {R_1R_2\over R_1+R_2} \bigr) u(d)
\label{der}
\end{equation}

\noindent where $u(d)$ is the potential energy between two parallel plates per unit area. The agreement between data and theory confirmed the existence of the Casimir force. 

Interestingly, in all modern experiments the surfaces were coated with a relatively thick metallic layer. This coat prevents the accumulation of electrostatic charges on the surfaces, which could seriously compromise the measurement. Furthermore, the use of thick films allows one to neglect possible size dependent effects such as those related to the {\it skin depth} (i.e., the penetration of the radiation inside the metallic layer), and to assume that the permittivities are those of the corresponding bulk metals, which can be easily found in literature. It is important to underline that, even in the case of simple geometries, it is necessary to know accurately the dielectric properties of the surfaces in order to compute the Casimir force with high precision.

The dependence of the Casimir force on the choice of the materials used in the experiment is a fascinating topic that has not received adequate attention. Vacuum fluctuations are so intimately connected to the dielectric function of the objects distributed in space that in principle one can accurately tailor the zero-point energy, and, therefore, the Casimir force, by engineering the boundary conditions for the electromagnetic field with a suitable choice of appropriately designed materials. 

As a first step in this new research direction, we have investigated the effect of Hydrogen Switchable Mirrors (HSMs) on the Casimir force. HSMs\cite{griessen1} are shiny metals in their {\it as deposited} state. However, when they are exposed to a hydrogen rich atmosphere, they become optically transparent. The effect is reversible. Since the fluctuations of the electromagnetic field depend on the optical properties of the surfaces, the attraction between two HSMs in air should be different from the attraction between the same HSMs immersed in a hydrogen rich atmosphere. In particular on intuitive grounds one expects that, in the transparent state, the Casimir force will be much weaker than in the reflective state.

We have measured the Casimir force between a gold-coated plate and a sphere coated with a HSM for separations in the $\simeq 70$ nm to $\simeq 400$ nm range. Although the switch from the reflective to the transparent state upon hydrogenation was confirmed in accordance with the literature, we did not observe any significant difference in the Casimir force measured in the two cases. This surprising and counterintuitive result can be explained using the Lifshitz theory\cite{lifshitz}, and sheds light on the role that optical wavelengths much larger than the separation between the surfaces play in the Casimir effect. 

\section{Experimental Apparatus}

\begin{figure}[t]
\epsfxsize=\textwidth
\epsfbox{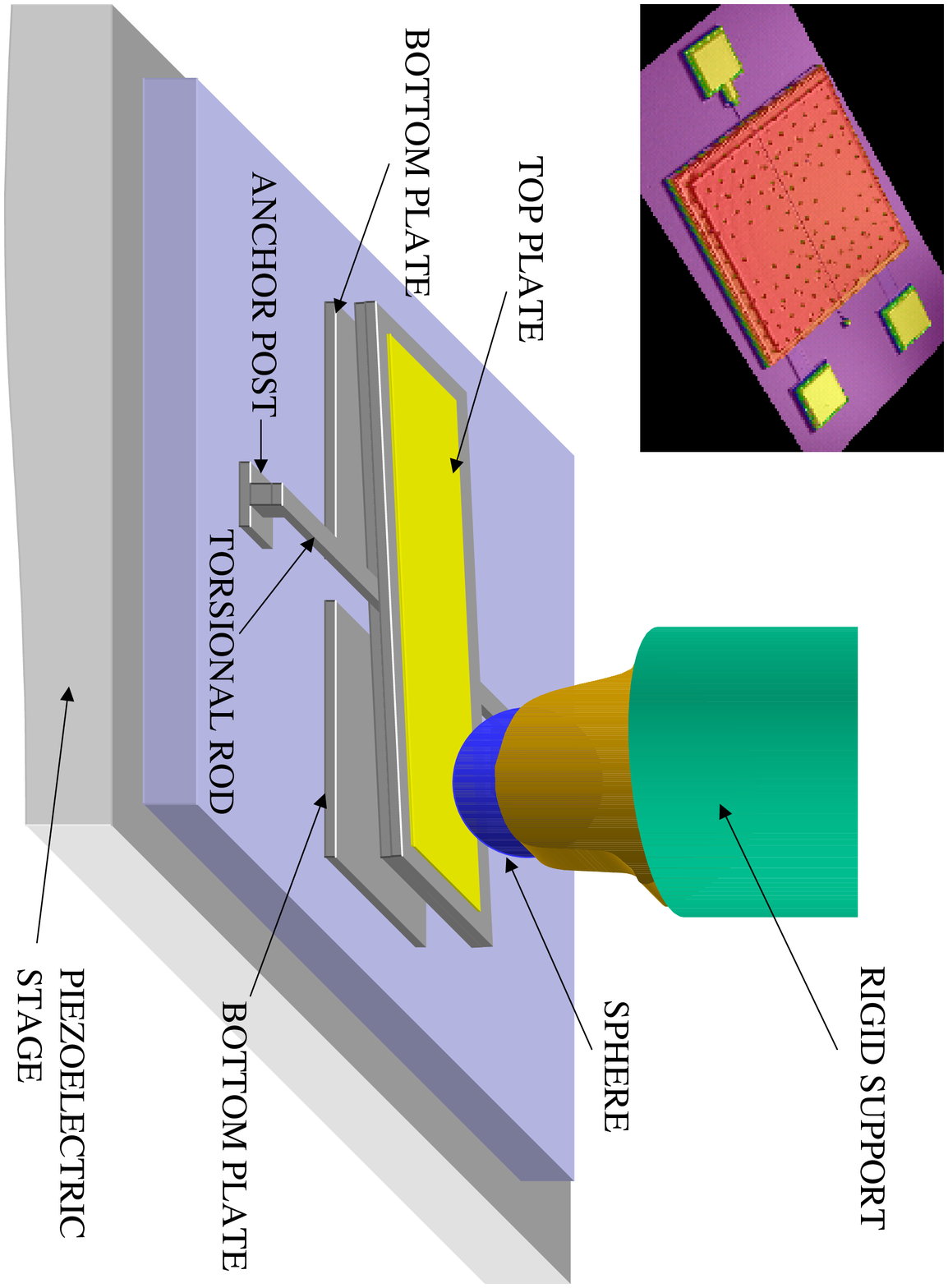}
\vspace{-30pt}
\caption{Sketch of the experimental apparatus. Inset: An image of the microtorsional device obtained with an optical profiler.\label{apparatus}}
\end{figure}

Our experimental apparatus\cite{hobun1,hobun2} is designed to study the Casimir effect between a sphere and a plate at sub-micron distances (see Fig. \ref{apparatus}). The measurement is carried out by bringing a metallized polystyrene sphere close to a freely suspended flat plate of a microtorsional device (MTD), and measuring the rotation angle of the plate induced by the Casimir attraction with the sphere. 

The MTD [fabricated by Cronos (acquired by MEMSCAP, Crolles, France) on the basis of our design, using MicroElectroMechanical Systems (MEMS) technology] consists of a gold-coated, 500 $\mu$m $\times$ 500 $\mu$m plate made of polysilicon and phosphosilicate glass that is suspended over the substrate with two thin polysilicon torsional rods (40 $\mu$m long, 2 $\mu$m thick and 2 $\mu$m wide). On the other end, the rods are anchored to the substrate by means of support posts. The top plate is thus free to rotate around the axis defined by the two rods. At 2 $\mu$m below the top plate, there are two 0.5 $\mu$m thick polysilicon electrodes, each of half the dimension of the top plate, located symmetrically with respect to the tilting axis. 

The capacitance between the top plate and each bottom electrode depends on the tilting angle $\theta$. If no force is applied to the MTD, the top plate is parallel to both electrodes, and the two capacitances are equal. As the top plate is tilted by an external force $F$, one of the two capacitances increases by $\delta C\propto \theta \propto F$, while the other decreases by the same amount. The top plate and the electrodes are connected to a capacitance bridge that allows very accurate measurements of $\delta C$\cite{hobun1}. AC excitations of opposite phase are applied to the two bottom plates, while the top plate is connected to a charge sensitive amplifier whose output is fed into a lock-in amplifier, locked at the frequency of the AC excitation (approximately $100$ kHz). This frequency is much higher than the mechanical resonance frequency of the MTD (approximately $2$ kHz), and does not induce any motion of the top plate. The amplitudes of the AC excitations are adjusted in such a way that, in the absence of an external torque, the output of the charge sensitive amplifier is zero. When the MTD is tilted by an external force, the bridge goes out of balance, and the output of the lock-in amplifier $A$ increases proportionally to $\delta C$. The apparatus is capable of measuring values of $\delta C$ in the $10^{-6}$ pF range, corresponding to rotations of the order of $10^{-7}$ rad. Since the spring constant of the MTD is of the order of $k_s=10^{-8}$ Nm/rad, the sensitivity in the torque measurement is $M=k_s\theta\simeq 10^{-15}$ Nm, corresponding to forces of the order of $F={M\over b}=10$ pN when the force is applied in the middle of one of the two arms of the MTD ($b=125$ $\mu$m), as in our experiment. Note that the proportionality constant $k$ that relates $A$ to $F$ is not known {\it a priori}, and must be determined by calibrating the apparatus with a known force. 

The MTD is mounted on the top of a calibrated piezoelectric mechanical translation stage (Polytec-PI, Auburn, MA). When a voltage is applied to the piezoelectric stage, the MTD is brought closer to the sphere by an amount $d_{pz}$ proportional to the applied voltage. The distance between the sphere and the plate is $d=d_0 - d_{pz}$\cite{noteford}, where $d_0$ is the value of $d$ when no voltage is applied to the piezoelectric stage. It is important to underline that, whereas $d_{pz}$ is known with a precision of less than 1 nm, $d_0$ is {\it a priori} unknown, and must also be determined during the calibration or with a proper analysis of the data.

A mechanical support keeps a $100$ $\mu$m radius polystyrene sphere (Duke Scientific) over the top plate of the MTD. The sphere is coated with a HSM. On the opposite side, the sphere is glued to a copper wire connected to a power supply. This allows us to apply a bias voltage between the sphere and the top plate of the MTD.

The history of HSMs dates back to 1996\cite{griessen1}, when it was observed that Pd-capped yttrium and lanthanum films can be switched from a shiny metal to an optically transparent semiconductor when exposed to a hydrogen atmosphere. This transition is induced by the formation of hydrides that alters the structure of the films. In order for the device to work, the film must be covered with a thin film of Pd (few tens of nanometers) that prevents the oxidation of the rare-earth elements and allows the dissociation of the hydrogen molecules. Similar effects have been observed recently in Pd-capped Mg-alloyed-rare-earth films\cite{nagengast} and Pd-capped Mg-X films, X being Ni, Co, Fe, or Mn\cite{richardson1,richardson2,isidorsson}. 

\begin{figure}[t]
\epsfxsize=\textwidth
\epsfbox{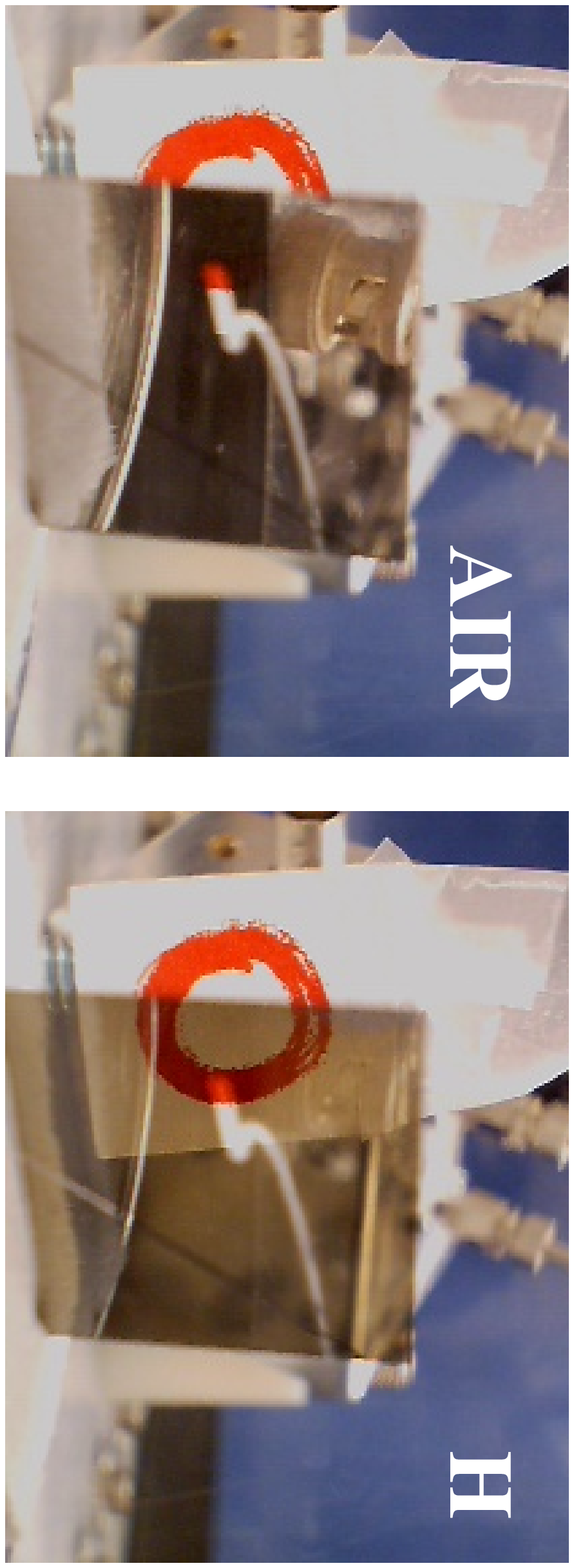}
\vspace{-30pt}
\caption{A Hydrogen Switchable Mirror in air and in hydrogen. A similar mirror was deposited on the sphere of our experimental apparatus.\label{film}}
\end{figure}

We have coated the sphere with a Pd-capped Mg-Ni HSM, obtained by repeating 7 consecutive evaporations of alternate layers of Mg ($100$ \AA) and Ni ($20$ \AA), followed by an evaporation of a thin film of Pd ($50$ \AA). In Fig. \ref{film}, we show a glass slide coated according to the procedure described, both in its {\it as deposited} state, and in its hydrogenated state. It is evident that the optical properties of the film are very different in the two situations. We have measured the transparency of the film over a wavelengths range between $0.5$ $\mu$m and $3$ $\mu$m, and its reflectivity at $\simeq 660$ nm, keeping the sample in air and in an argon-hydrogen atmosphere ($4\%$ H$_2$ vol/vol). The results are in good agreement with the values reported in \cite{richardson1}.

\section{Calibration and Measurement Technique}

In all modern experiments on the Casimir force, the determination of the initial distance $d_0$ between the two interacting surfaces is one of the main sources of error (for a discussion of this topic, see for example \cite{iannuzzi}). The easiest way to solve this problem is to fit directly the Casimir force data with the theoretical result, keeping $d_0$ as a free parameter\cite{hobun1,ederth}. In our case, since the dielectric function of HSMs is known only in a limited range of frequencies, we cannot calculate the theoretical curve, and, therefore, we cannot use this procedure. 

Alternatively, it is possible to determine $d_0$ by applying a bias between the two surfaces and by measuring the electrostatic attraction as a function of the distance. This also allows one to calibrate the force sensor, i.e. to measure the proportional constant $k$ that relates the output of the read-out electronics to the force between the two interacting surfaces. 
Our attempts to use this method were not successful. The results were not reproducible and different calibration usually gave slightly different values of $d_0$. In our opinion, the method is not very precise, especially if the measurements are taken at very short distances and at atmospheric pressure, as in our case.

To solve the $d_0$ problem, we have implemented a technique in which the calibration and the measurement of the Casimir force are performed simultaneously.

The force between the sphere and the plate when a bias voltage $V_{bias}$ is applied between the two surfaces is given by the sum of the electrostatic and the Casimir attractions:

\begin{equation}
|F|=\epsilon_0 \pi R {(V_{bias}+V_0)^2\over (d_0-d_{pz})}+|F_C|
\label{force}
\end{equation}

\noindent where $\epsilon_0$ is the permittivity of vacuum. The {\it residual voltage} $V_0$ arises from the difference of the coating film work-functions\cite{over}. It is evident that a measurement of $F$ when $V_{bias}=-V_0$ gives the value of $F_C$ directly. 

In order to measure $k$, $d_0$, $V_0$, and $F_C$ simultaneously, we note that for a given value of $d_{pz}$, the output of the lock-in amplifier $A$ is a quadratic function of $V_{bias}$. From equation \ref{force}, in fact, one obtains:

\begin{equation}
A=k\epsilon_0 \pi R {(V_{bias}+V_0)^2\over (d_0-d_{pz})}+k|F_C|
\label{quadra}
\end{equation}

\begin{figure}[t]
\epsfxsize=\textwidth
\epsfbox{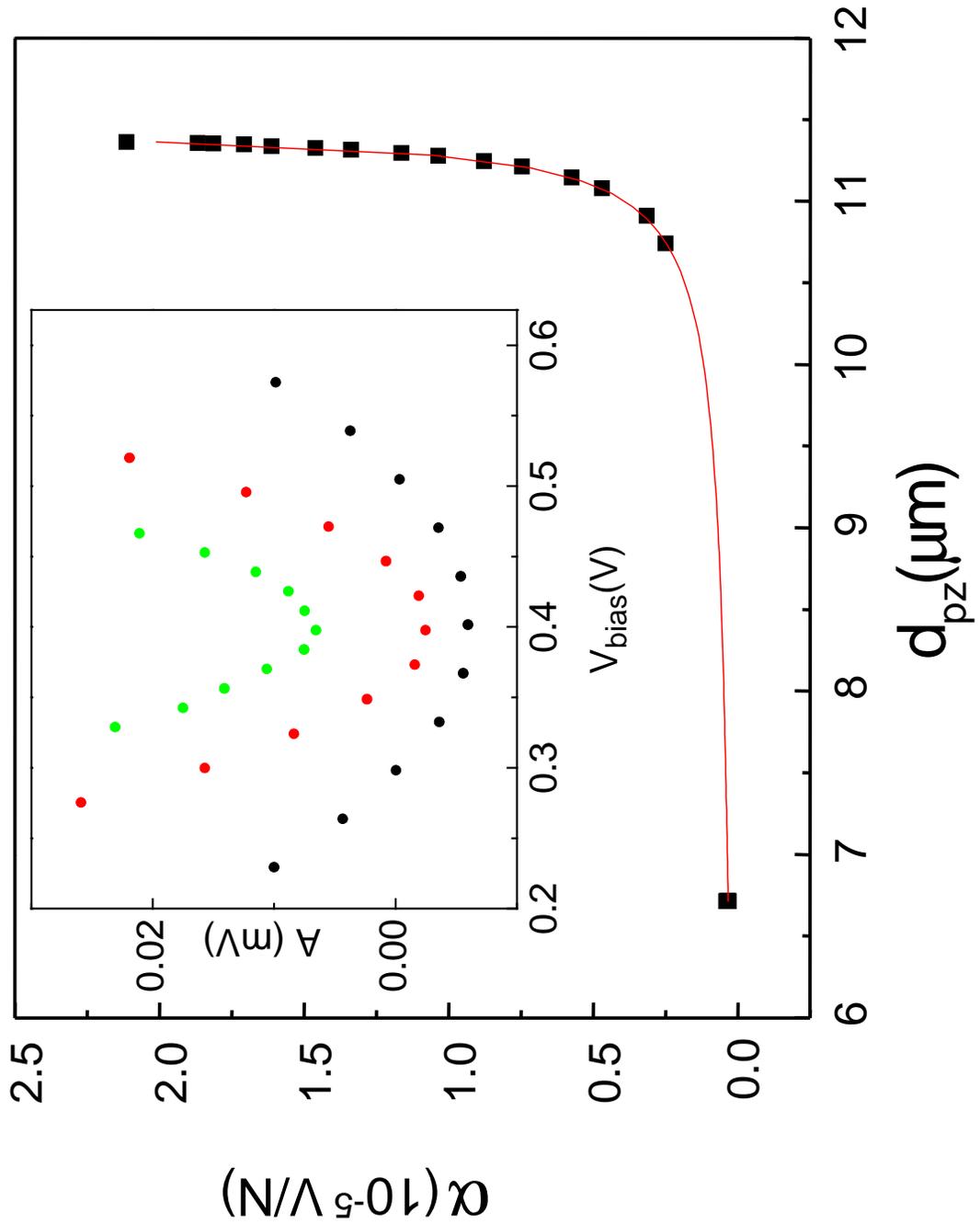}
\vspace{-30pt}
\caption{Determination of the parameter $k$ and $d_0$ through the fit of $\alpha$ as a function of $d_{pz}$ (see text). Inset: three parabolas obtained by scanning the voltage on the sphere ($V_{bias}$) for three different values of $d_{pz}$. $A$ is the output of the lock-in amplifier.\label{calibration}}
\end{figure}

\noindent In our experimental procedure, we measure $A$ as a function of $V_{bias}$, and we repeat this measurement for several values of $d_{pz}$. For example, in the inset of Fig. \ref{calibration} we show three of these measurements obtained for three different values of $d_{pz}$. As expected, the experimental data are distributed along a parabola. Then, for each $d_{pz}$, we fit these parabolas using a generic quadratic equation $y=\alpha(x+x_0)^2+\beta$, where $\alpha$, $\beta$, and $x_0$ are free parameters. We note that:

\begin{equation}
\alpha=k{\epsilon_0 \pi R \over (d_0-d_{pz})}
\label{coeff}
\end{equation}

\noindent We can thus determine $k$ and $d_0$ by fitting $\alpha$ as a function of $d_{pz}$ with the function in equation \ref{coeff}. In Fig. \ref{calibration} we show the result of this interpolation. Once $k$ is known, $F_C$ can be determined for all the value of $d_{pz}$ by means of:

\begin{equation}
|F_C|={\beta \over k}
\label{cf}
\end{equation}

\noindent Since we have already evaluated $d_0$, we can finally plot $F_C$ as a function of $d=d_0-d_{pz}$.

\section{Results and Discussion}

\begin{figure}[t]
\epsfxsize=\textwidth
\epsfbox{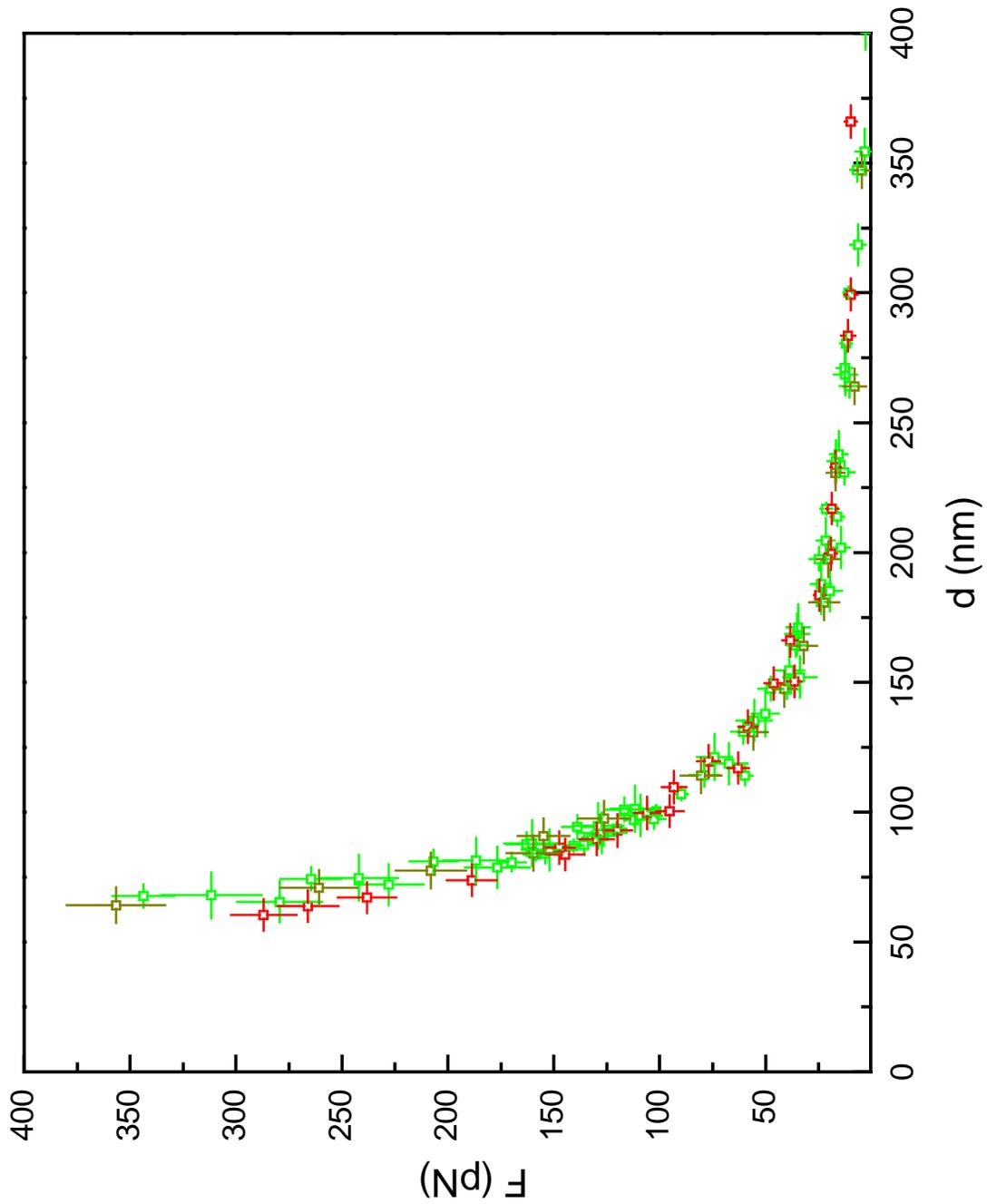}
\vspace{-30pt}
\caption{Casimir force between a gold-coated plate and a sphere coated with a Hydrogen Switchable mirror as a function of the distance, in air (green dots) and in argon-hydrogen (red dots).
\label{global}}
\end{figure}

Using one of the spheres coated with the procedure described in section 2, we carried out the measurement of the Casimir force in air. Then we filled the experimental apparatus with an argon-hydrogen gaseous mixtures, and, after few hours, we measured the force again. Both the measurements were also repeated using another similar sphere and a different MTD. In total, we carried out the measurement 5 times in air and 3 times in argon-hydrogen. The results are shown in Fig. \ref{global}. The error bars on the force are the result of the propagation of the errors on $\beta$ and $k$ resulting, respectively, from the fit of the parabolas and of the calibration curve (equation \ref{coeff}). The error bars on the distance is the error on $d_0$ also obtained from the calibration. It is evident that the force does not change significantly upon hydrogenation of the HSM\cite{secmes}.

\begin{figure}[t]
\epsfxsize=\textwidth
\epsfbox{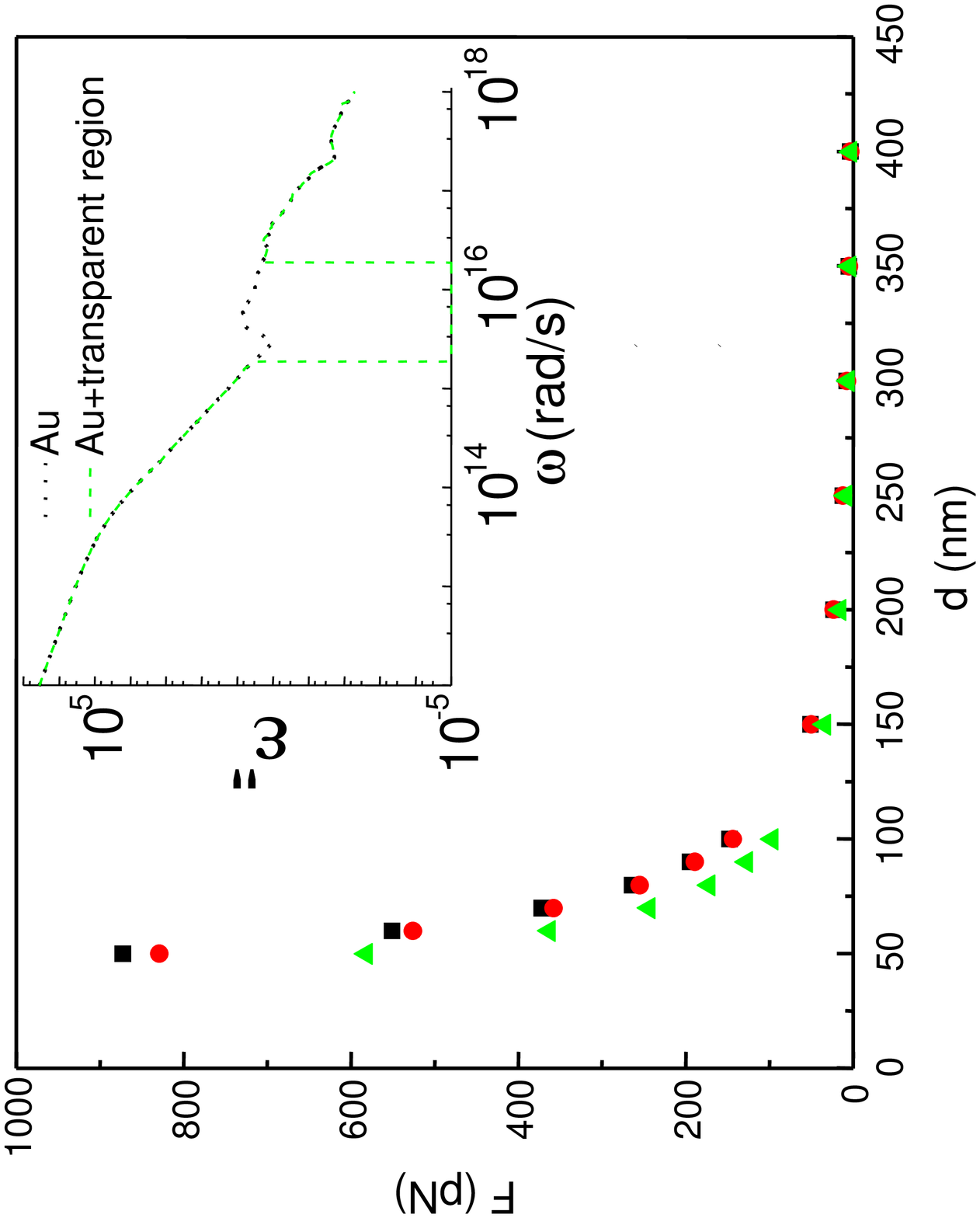}
\vspace{-30pt}
\caption{Results of the calculation of the Casimir force using the Lifshitz formula for a material with dielectric function equal to that of gold with exception of a frequency range where the material is supposed to be transparent (see inset). Black squares: pure gold; red circles: transparent window from $0.2$ $\mu$m to $2.5$ $\mu$m; green triangles: transparent window from $1$ $\mu$m to $200$ $\mu$m.\label{calculus}}
\end{figure}

In order to explain this counterintuitive result, we first note that the dielectric properties of the HSMs used in this experiment are known only in a limited range of wavelengths $\lambda$, spanning from approximately $0.3$ $\mu$m to $2.5$ $\mu$m\cite{richardson1}. However, since the separation between the sphere and the plate in our experiment is in the $\simeq 100$ nm range, one expects that it is not necessary to know the dielectric function for $\lambda>>2.5$ $\mu$m, because those modes should not give rise to large contributions to the force. We have thus performed a mathematical exercise to see if this intuitive argument is correct. The Casimir attraction in vacuum between a sphere of radius $R$ and a plate with dielectric function $\epsilon(\omega)$ is given by\cite{lifshitz}:

\begin{equation}
\begin{split}
F=&{\hbar R \over 2\pi c^2}\int_0^\infty \int_1^\infty p \xi ^2 \\ &\left\{ \ln \left[ 1-{(s-p)^2\over (s+p)^2}e^{-2pz\xi\over c}\right] + \ln \left[ 1-{(s-p\epsilon)^2\over (s+p\epsilon)^2}e^{-2pz\xi\over c}\right] \right\}dpd\xi
\end{split}
\label{lif}
\end{equation}

\noindent where $s=\sqrt{\epsilon-1+p^2}$, $\epsilon$ is the dielectric function of the material of the two surfaces calculated at imaginary frequency $i\xi$, and $c$ and $\hbar$ are the usual fundamental constants. Using equation \ref{lif}, we have calculated the Casimir force for a material with hypothetical dielectric function equal to the dielectric function of gold at all $\lambda$, except for a wavelength interval spanning from $\lambda_{min}=0.2$ $\mu$m to $\lambda_{max}=2.5$ $\mu$m (see inset of Fig. \ref{calculus}), where we have supposed the material to be completely transparent. This procedure mimics the effect of hydrogenation. The result, compared to the theoretical force expected for two surfaces made of gold, is shown in Fig. \ref{calculus}. The presence of a transparency window in the visible and near-infrared wavelength range does not substantially alter the Casimir attraction. In order to have large variations of the Casimir force, it is necessary to have a much wider transparency window, such as for example from $\lambda=1$ $\mu$m to $\lambda=200$ $\mu$m, as shown in Fig. \ref{calculus}. If we assume that the HSM layers deposited on the sphere change their dielectric properties only in a limited wavelength range, which is not in contradiction with the experimental data available in literature, then the decrease of the force upon hydrogenation could be too small to be observable.

It is interesting to underline that the dependence of the force from the dielectric properties of the materials, as described by equation \ref{lif}, is mathematically connected to the dielectric function calculated on the imaginary axis of frequencies, which can be determined from the equation\cite{repmoh}: 

\begin{equation}
\epsilon(i\xi)=1+{2\over \pi}\int_0^\infty {{\omega \epsilon''(\omega) \over \omega^2 + \xi^2} d\omega}
\label{epsilon}
\end{equation}

\noindent where $\epsilon''$ is the imaginary part of the dielectric function. The integral in equation \ref{epsilon} runs over all real frequencies, with non-negligible contributions arising from a very wide range of frequencies, as already noted in \cite{bo}. This explains the results illustrated in Fig. \ref{calculus}.

Although the arguments reported above are very convincing, we want to underline that, in our discussion, we have not accounted for the presence of the thin Pd layer on top of the HSM. The thickness of this layer, in fact, is of the order of $50$ \AA. The skin depth of ultraviolet, visible, and infrared radiation, is of the order of $100$ \AA. Therefore, the Pd layer can probably be considered transparent at all the wavelengths relevant to the calculation of the Casimir force. Nevertheless, a pioneering paper by S. L. Tan and P. W. Anderson\cite{ande} showed that, if the dispersion relation of two-dimensional plasmons is included in the calculation of the non-retarded Van der Waals forces between low-dimensional metals such as graphite or polyacetylene, one obtains a dependence on distance $d$ significantly different from the ${1\over d^3}$ theoretical prediction. Similar effects cannot be ruled out {\it a priori} in the case of the retarded Casimir force between ultrathin metallic films. Further experiments are needed to understand the role of the thin layer of Pd in the Casimir attraction between HSMs. 

\section{Conclusions}

We have presented the first attempt to tune the Casimir force with engineered materials. In particular, we have measured the attraction between a gold coated plate and a sphere coated with a HSM in air and in a hydrogen rich atmosphere. Intuitively, a large change of the force is expected upon hydrogenation. On the contrary, no significant difference was observed. This result can be explained assuming that the HSMs used in the experiment switch only in correspondence of a limited wavelength region. 

\section{Acknowledgements}

One of the author (D.I.) is especially indebted with H. B. Chan for his initial technical support. R. Griessen, B. Dam, G. Kowach, Y. Barash, and C. Gmachl are also acknowledged for assistance and useful discussions. The initial part of this work was performed at Bell Labs, Lucent Technologies. This project was partially supported by the National Science Foundation under the Grant No. PHY-0117795.

\newpage

\newpage

\newpage

\newpage

\newpage

\begin{thebibliography}{99}

\bibitem{milonni} For a general introduction to quantum electrodynamics, see for example P. W. Milonni (1993) {\it The Quantum Vacuum:An Introduction to Quantum Electrodynamics} (Academic Press, San Diego)

\bibitem{casimir} H. B. G. Casimir (1948) {\it Proc. K. Ned. Akad. Wet.}  {\bf 51}, 193-195 

\bibitem{serry} F. M. Serry, D. Walliser, and G. J. Maclay (1995) {\it J. Microelectromech. Syst.} {\bf 4}, 193-205 

\bibitem{hobun1} H. B. Chan, V. A. Aksyuk, R. N. Kleinman, D. J. Bishop, and F. Capasso (2001), {\it Science} {\bf 291}, 1941-1944

\bibitem{hobun2} H. B. Chan, V. A. Aksyuk, R. N. Kleinman, D. J. Bishop, and F. Capasso (2001), {\it Phys. Rev. Lett.}  {\bf 87}, 260402 

\bibitem{lifshitz} E. M. Lifshitz (1956), {\it Sov. Phys. JETP} {\bf 2}, 73-83 

\bibitem{over} P. H. G. M. van Blokland and J. T. G. Overbeek (1978) {\it J. Chem. Soc. Far. Trans.} {\bf 24}, 2637-2651

\bibitem{lamo} S. K. Lamoreaux (1997) {\it Phys. Rev. Lett.} {\bf 78}, 5-8

\bibitem{moh1} U. Mohideen, A. Roy (1998) {\it Phys. Rev. Lett.} {\bf 81}, 4549-4552

\bibitem{moh2} A. Roy, C.-Y. Lin, U. Mohideen (1999) {\it Phys. Rev.} {\bf D60}, 111101

\bibitem{moh3} B. W. Harris, F. Chen, U. Mohideen (2000) {\it Phys. Rev.} {\bf A62}, 052109

\bibitem{ederth} T. Ederth, {\it Phys. Rev.} {\bf A62}, 062104

\bibitem{bressi} G. Bressi, G. Carugno, R. Onofrio, and P. Ruoso (2002), {\it Phys. Rev. Lett.} {\bf 88}, 041804 

\bibitem{decca1} R. Decca, D. Lopez, E. Fischbach, and D. E. Krause (2003), {\it Phys. Rev. Lett.} {\bf 91}, 050402

\bibitem{decca2} R. S. Decca, E. Fischbach, G. L. Klimchitskaya, D. E. Krause, D. Lopez, and V. M. Mostepanenko (2003), {\it Phys. Rev.} {\bf D68}, 116003

\bibitem{repmoh} M. Bordag, U. Mohideen, and V. M. Mostepanenko (2001), {\it Phys. Rep.} {\bf 353}, 1-205

\bibitem{israe} J. N. Israelachvili (1991) {\it Intermolecular and Surface Forces} (Academic Press, London)

\bibitem{iannuzzi} D. Iannuzzi, I. Gelfand, M. Lisanti, and F. Capasso (2004), {\it Proceedings of the 6th Workshop on Quantum Field Theory Under the Influence of External Conditions} (Rinton Press) {\it in press}

\bibitem{newmost} F. Chen, G. L. Klimchitskaya, U. Mohideen, and V. M. Mostepanenko (2004), {\it Phys. Rev.} {\bf A} {\it in press}

\bibitem{griessen1} J.N. Huiberts, R. Griessen, J. H. Rector, R. J. Wijngaarden, J. P. Dekker, D. G. de Groot, and N. J. Koeman. (1996), Nature {\bf 380}, 231-234 

\bibitem{noteford} Rigorously, one should also consider that the rotation induced by the force on the oscillator reduces the distance between the sphere and the plate. However, since we are interested in comparing two different measurements (the measurement in air and the measurement in hydrogen rich atmosphere), this systematic error can be neglected.

\bibitem{nagengast} D. G. Nagengast, A. T. M. Van Gogh, E. S. Kooij, B. Dam, and R. Griessen (1999) {\it Appl. Phys. Lett.} {\bf 75}, 2050-2053

\bibitem{richardson1} T. J. Richardson, J. L Slack, R. D. Armitage, R. Kostecki, B. Farangis, and M. D. Rubin (2001), {\it Appl. Phys. Lett.} {\bf 78}, 3047-3049

\bibitem{richardson2} T. J. Richardson, J. L. Slack, B. Farangis, and M. D. Rubin (2002), {\it Appl. Phys. Lett.} {\bf 80}, 1349-1351

\bibitem{isidorsson} J. Isidorsson, I. A. M. E. Giebels, R. Griessen, and M. Di Vece (2002), {\it Appl. Phys. Lett.} {\bf 80}, 2305-2307

\bibitem{secmes} We have also carried out some measurements using a third sphere. In that case, we observed that in air the Casimir force at short distances was larger compared to the result shown in Fig. \ref{global}. We attribute this discrepany to the larger roughness of the film deposited on this sphere. However, also in that case, the Casimir force did not change significantly upon hydrogenation.   

\bibitem{bo} M. Bostr\"{o}m and B. E. Sernelius (1999), {\it Phys. Rev.} {\bf B61}, 2204-2210

\bibitem{ande} S.L. Tan and P.W. Anderson (1983), {\it Chem. Phys. Lett.} {\bf 97}, 23-25.



\end{thebibliography}
\end{document}